\documentclass{article}

\usepackage{arxiv}

\usepackage[utf8]{inputenc} 
\usepackage[T1]{fontenc}    
\usepackage{hyperref}       
\usepackage{url}            
\usepackage{booktabs}       
\usepackage{amsfonts}       
\usepackage{nicefrac}       
\usepackage{microtype}      
\usepackage{lipsum}		
\usepackage{graphicx}
\usepackage{doi}

\title{Mighty Tracker - Performance Studies of the MightyPix for LHCb}

\author{ {Hannah Schmitz} \\
	HISKP Bonn\\
	University of Bonn\\
	   53115 Bonn, Germany \\
	\texttt{hannah.schmitz@cern.ch} \\
	\\
	\And
	{Lucas Dittmann} \\
	Physikalisches Institut\\
	University of  Heidelberg\\
	69120 Heidelberg, Germany \\
	 \texttt{lucas.marvin.dittmann@cern.ch} \\
	 \AND
	{Klaas Padeken} \\
	HISKP Bonn\\
	University of Bonn\\
	   53115 Bonn, Germany \\
	\texttt{klaas.padeken@cern.ch} \\
	\\
	\And
 {Sebastian Neubert} \\
	HISKP Bonn\\
	University of Bonn\\
	   53115 Bonn, Germany \\
	\texttt{sebastian.neubert@cern.ch} \\
	\\
}

\renewcommand{\headeright}{}
\renewcommand{\undertitle}{Prepared for submission to NIMA}

\hypersetup{
pdftitle={Mighty Tracker - Performance Studies of the MightyPix for LHCb},
pdfsubject={q-bio.NC, q-bio.QM},
pdfauthor={H.~Schmitz, L.~Dittmann, K.~Padeken, S.~Neubert},
pdfkeywords={First keyword, Second keyword, More},
}

\begin{document}
\maketitle

\begin{abstract}
	A new downstream tracking system, known as the Mighty Tracker, is planned to be installed at LHCb during LS4 of the LHC. This will allow an increase in instantaneous luminosity from $2\cdot10^{33}~\mathrm{cm}^{-2}\mathrm{s}^{-1}$ to $1.5\cdot10^{34}~\mathrm{cm}^{-2}\mathrm{s}^{-1}$ and therefore an overall higher irradiation and up to six times higher occupancy.
	To keep the material budget as low or even lower as for the current detector, the Mighty Tracker is planned as a hybrid system with silicon pixels in the inner and scintillating fibers in the outer region.\newline
	For the pixel detector part HV-CMOS MAPS with a pixel size of  
 $55~\mathrm{x}~165~\mathrm{\mu m^2}$ will be used. This technology has been chosen because of its low production costs, low material budget, high radiation tolerance and good timing resolution. 
	To fulfill the requirements, the development and characterization of the sensors focuses on radiation damage with fluences up to $2\cdot10^{15~}\mathrm{MeVn_{eq}}/\mathrm{cm^2}$ and a timing resolution $\leq3~\mathrm{ns}$.
	The timing resolution is restricted due to the $40~\mathrm{MHz}$ trigger-less DAQ by LHCb.\newline
	To characterize and further develop the MightyPix a new readout system called MARS~(Mighty TrAcker Readout System) has been developed.
	 MARS is a modular system, able to test different single chips in the laboratory as well as at testbeam facilities.
	It has been used to perform first characterization studies of development-chips investigating radiation tolerance and timing resolution as well as the dependence of the sensor settings on the overall performance.
\end{abstract}

\keywords{Large Scale Pixel Detector \and HVCMOS MAPS \and Readout System Development \and LHCb Upgrade}

\section{Introduction}
\label{sec:intro}

The LHCb experiment is located at point $8$ of the LHC at CERN. Due to the design of the LHCb detector as a single-arm forward spectrometer flavor physics as well as searches for new physics beyond the Standard Model can be performed with unprecedented accuracy~\cite{CERN-LHCC-2021-012}.
The next big step to further increase sensitivity for new phenomena will be to collect more data. Thus, the instantaneous luminosity will increase from  $2\cdot10^{33}~\mathrm{cm}^{-2}\mathrm{s}^{-1}$ to $1.5\cdot10^{34}~\mathrm{cm}^{-2}\mathrm{s}^{-1}$ by the start of Run $5$. Because of that, approximately $300~\mathrm{fb}^{-1}$ will be collected by the end of Run $6$~\cite{CERN-LHCC-2021-012}. To provide precise tracking and particle identification, the overall detector has to be improved during Long Shutdown $4$. These improvements are called Upgrade $2$. 

All subdectectors will have to cope the increased occupancy and radiation levels. One key component are the tracking stations downstream of the dipole magnet - currently equipped with scintillating fibers. Those fibers cannot cope with the upcoming requirements in the high $\eta$ region around the beampipe and thus have to be replaced. Therefore, a hybrid detector consisting of silicon pixels in the high occupancy region and scintillating fibers in the outermost region is foreseen. This detector is called Mighty Tracker.\newline
The requirements and developments of the silicon pixel detector - MightyPix - are discussed in the following.

\section{MightyPix Requirements}
\label{sec:MightyPix_requirements}

\begin{figure}[htb]
    \centering
    \includegraphics[width =0.7\linewidth]{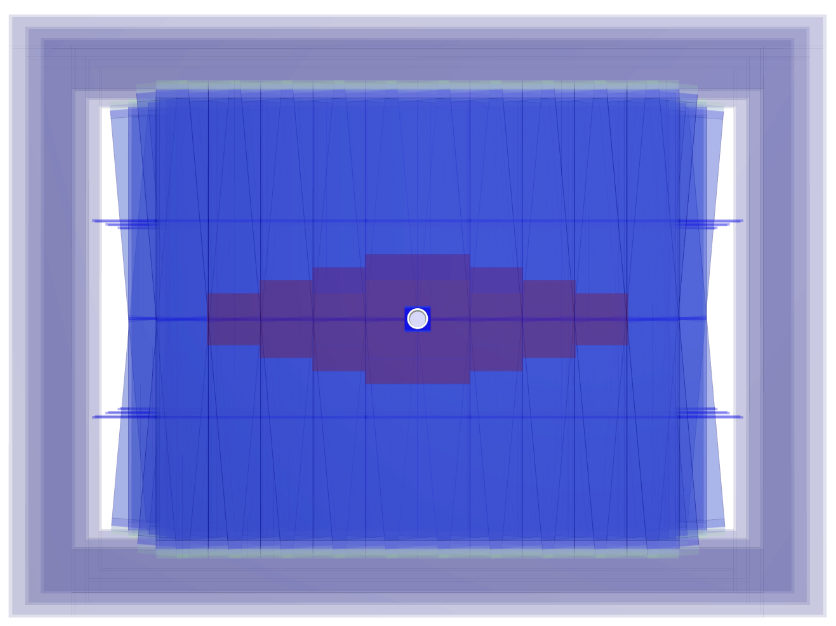}
    \caption{One Mighty Tracker module consisting of scintillating fibers, colored in blue, and pixel modules, colored in purple. The module is held by a C frame~(grey)~\cite{TaiHua}.}
    \label{fig:MightyTracker_Module}
\end{figure}

Due to the increased luminosity the Mighty Tracker has to withstand a radiation up to  $2\cdot10^{15}~\mathrm{MeV}\,\mathrm{n_{eq}}/\mathrm{cm^2}$. This is an increase by three orders in magnitude~\cite{KIRN2017481}. In addition, a time resolution $\leq3~\mathrm{ns}$ is required by the $40~\mathrm{MHz}$ trigger-less DAQ system of LHCb which was installed during LS3~\cite{CERN-LHCC-2021-012} and reads out the full detector at the LHC bunch crossing frequency. Thus, the collected data is fully reconstructed to further analysis processes without reduction by a hardware trigger. Therefore, a data output rate of $5~\mathrm{Gb}$ per chip is required. Besides these requirements, the power consumption has to be below $150~\mathrm{mW}$ and the material budget $\mathrm{X}_0\leq 2~\%$ per tracking layer. Those requirements are given by the location downstream of the magnet and within the inner part of the fiber tracker. Thus, as less material budget as possible is added to the overall hybrid detector. It's design is shown in figure~\ref{fig:MightyTracker_Module}. The pixel part consists of $26$ long and $2$ short submodules per layer, colored in purple. Each submodule is equipped with $35$ MightyPix sensors of size $52.8~\mathrm{x}~20~\mathrm{cm}$~\cite{CERN-LHCC-2021-012}. In total six layers are planned to be installed. The Mighty Tracker will consist of $18~\mathrm{m^2}$ pixel detector.

\begin{figure}[htb]
    \centering
    \includegraphics[width =0.7\linewidth]{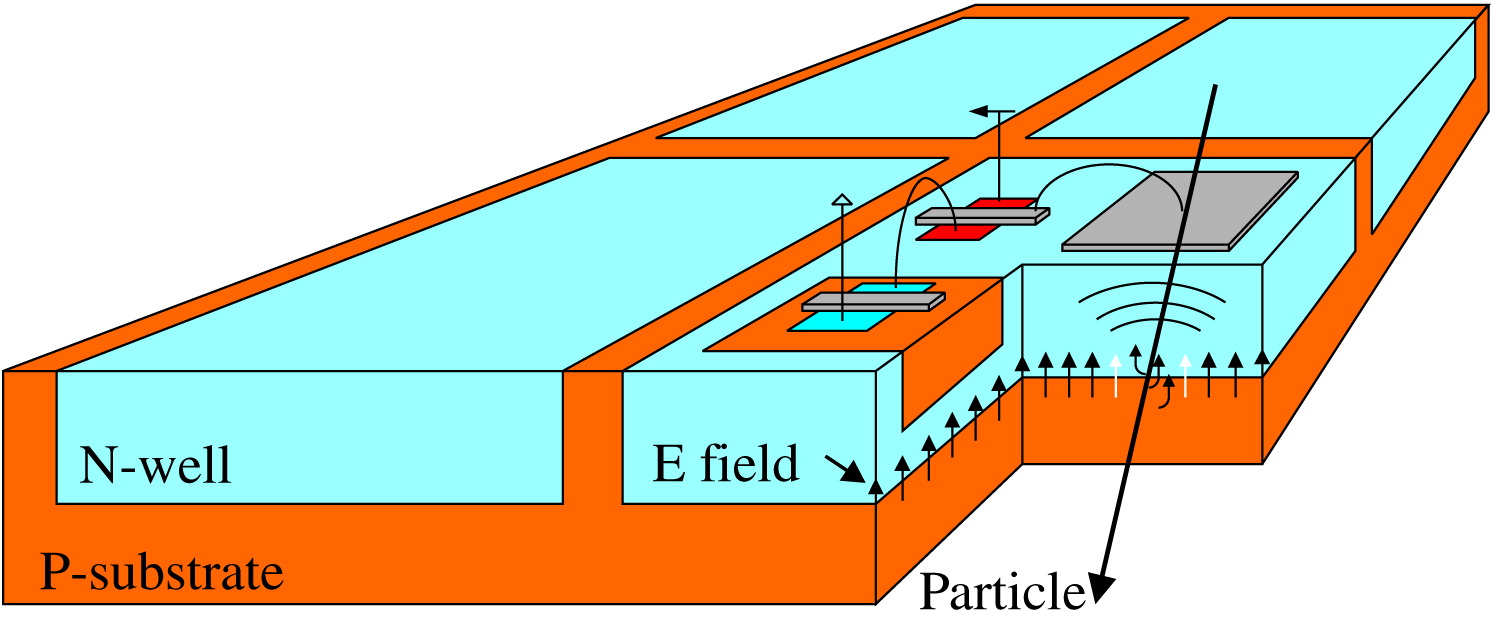}
    \caption{Monolithic detector consisting of four blocks CMOS technology~\cite{4437188}.}
    \label{fig:HVCMOS}
\end{figure}

To fulfill all requirements, HVCMOS MAPS are foreseen for the pixel sensors. The design of the sensor is done at KIT~\cite{4437188}. A schematic of a monolithic detector consisting of four blocks is shown in figure~\ref{fig:HVCMOS}.

Due to the integration of the readout electronics into the active sensor the material budget of the detector is low by design. In addition, a fast signal collection by drift is given as well as intrinsic radiation hardness due to the large collection electrode. 
Furthermore, a high breakdown voltage up to $200~\mathrm{V}$~\cite{Hirono:2876027} can be achieved with this technology.

\section{MightyPix Developments}
\label{sec:MightyPix_developments}

A first version of the MightyPix sensor was produced in a $180~\mathrm{nm}$ TSI process. The sensor has a size of $5~\mathrm{x}~20~\mathrm{mm}$ which is a quarter in $x$ of the final sensor size and a pixel size of $55~\mathrm{x}~165~\mathrm{\mu m^2}$. Its matrix consists of $29$ columns and $320$ rows~\cite{Hirono:2876027}. 

The analogue design of the sensor is based on other sensors of the HV-CMOS family as MuPix8, ATLASPix3.0, ATLASPix3.1 and TelePix~\cite{Peric:2021bcu}~\cite{Augustin:2020pkv}. Additionally, different engineering sensors called Run2020 were produced to test different pixel sizes as well as chip electronics focusing on the pixel amplifier and comparator~\cite{run2020}. Accordingly, the influence of different combinations of PMOS-, NMOS- and CMOS-type electronics on the sensor performance were studied.  

From laboratory measurements as well as a testbeam campaign at DESY in 2022 where the ATLASPix3.1 was characterized it was concluded that CMOS electronics for the comparater and the pixel amplifier are essential to achieve the requirements in terms of radiation hardness and time resolution~\cite{Padeken:2867083}~\cite{Padeken:2871825}. Besides studies on the dependence of the time resolution and the radiation hardness, the dependence of the overall sensor performance and the operating temperature at different radiation doses was studied. From that an increased performance~(efficiency $\epsilon\geq 99~\%$) at low temperatures~($-10~\mathrm{^\circ C}$) over a large pixel threshold range could be concluded. Similar studies are planned for the MightyPix to measure the influence of the improved pixel design on the overall performance.

In contrast, to the other sensors the digital part of the MightyPix will be compatible with the LHCb readout system. Thus, slow control~(I2C communication) as well as fast and time control~(TFC) are implemented. TFC operates at $320~\mathrm{MHz}$ which is  provided internally~\cite{Hirono:2876027}. 
The data link speed is $4~\mathrm{x}~1.28~\mathrm{Gbps}$ for a data output rate of $40~\mathrm{MHz}$ of the $32~\mathrm{bit}$ hit word. Besides this two slower modes can be chosen for the data output rate of the final state machine~($20~\mathrm{MHz}$ and $10~\mathrm{MHz}$). Hence, the data link speed is reduced to $640~\mathrm{MHz}$ resp. $320~\mathrm{MHz}$~\cite{Hirono:2876027}.

\section{Development of MARS}
\label{sec:MARS}

To characterize the sensors, a new readout system was developed. Previously used systems~(GECCO~\cite{gecco} and MUDAQ) are still in use but cannot be used optimally for the planned measurements due to their design. For example, GECCO cannot be placed easily in a testbeam telescope. MUDAQ, as well as GECCO, is a long-term established readout system but new systems cannot simply be rebuilt as some parts are no longer available.
Therefore, the main design criteria of MARS are long-term usability as well as the usage as a single-sensor laboratory test setup while providing the flexibility to operate the system as part of a beam telescope. Additionally, it was designed to easily adapt the system to different devices. So far the system is set up for Run2020, TelePix and MightyPix sensors. Moreover, chip inserts from other systems can be used by an additional adapter. Hence, one sensor can be characterized with two different setups to exclude systematic uncertainties originating by hardware. 

Another key element of the design is the possibility to operate the sensor at its design speed. Thus, high data output rates up to $1.28~\mathrm{GHz}$ have to be processed. Therefore, the firmware and hardware was designed to enable frequencies up to $1.6~\mathrm{GHz}$.

\begin{figure}[htbp]
    \centering
    \includegraphics[width=0.9\linewidth]{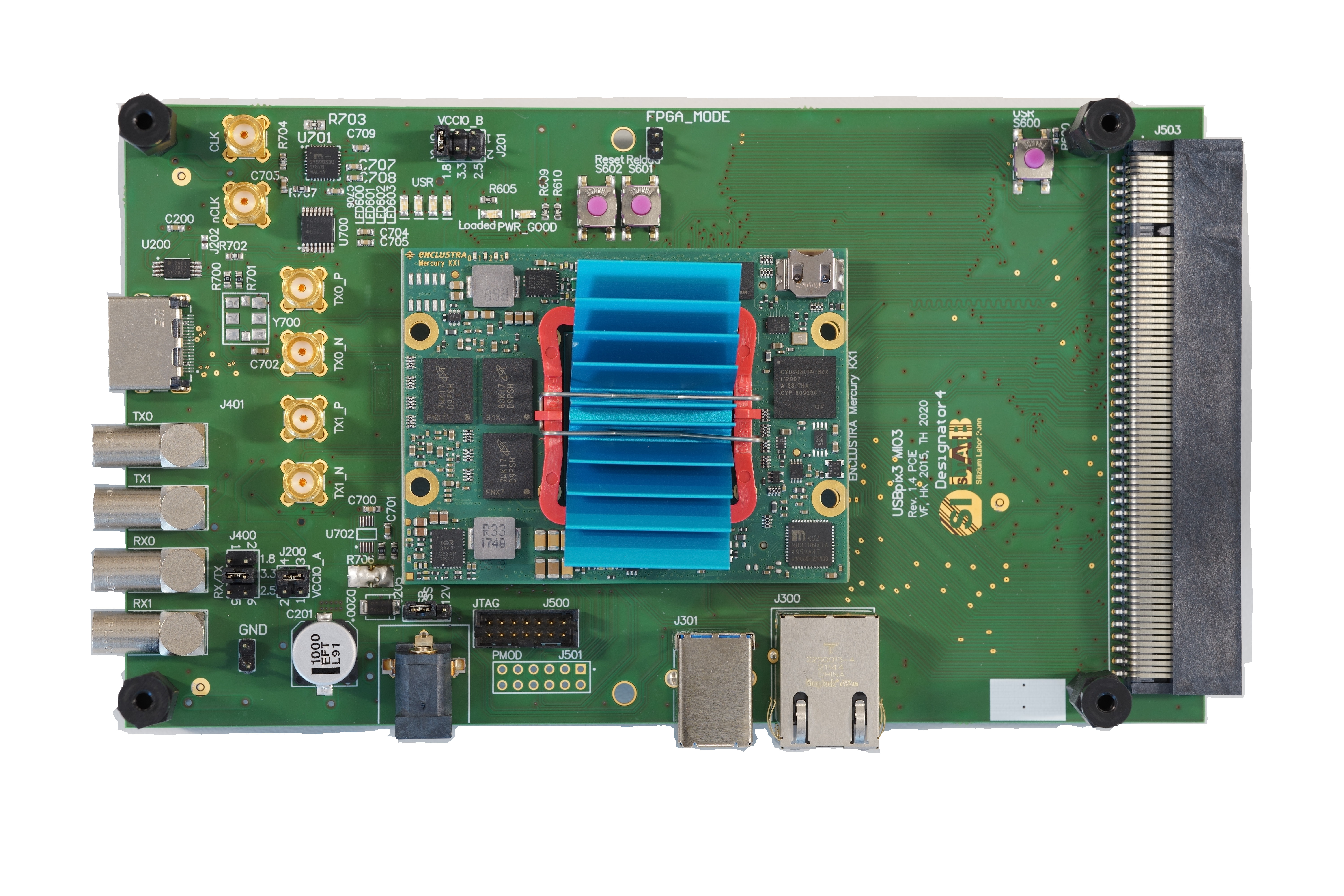}
    \caption{FPGA covered by a blue heat sink placed on the FPGA carrier board equipped with all components and connectors for the adapter board as well as the back-end electronics.}
    \label{fig:FPGABoard}
\end{figure}

\begin{figure}[htbp]
    \centering
    \includegraphics[width=0.8\linewidth]{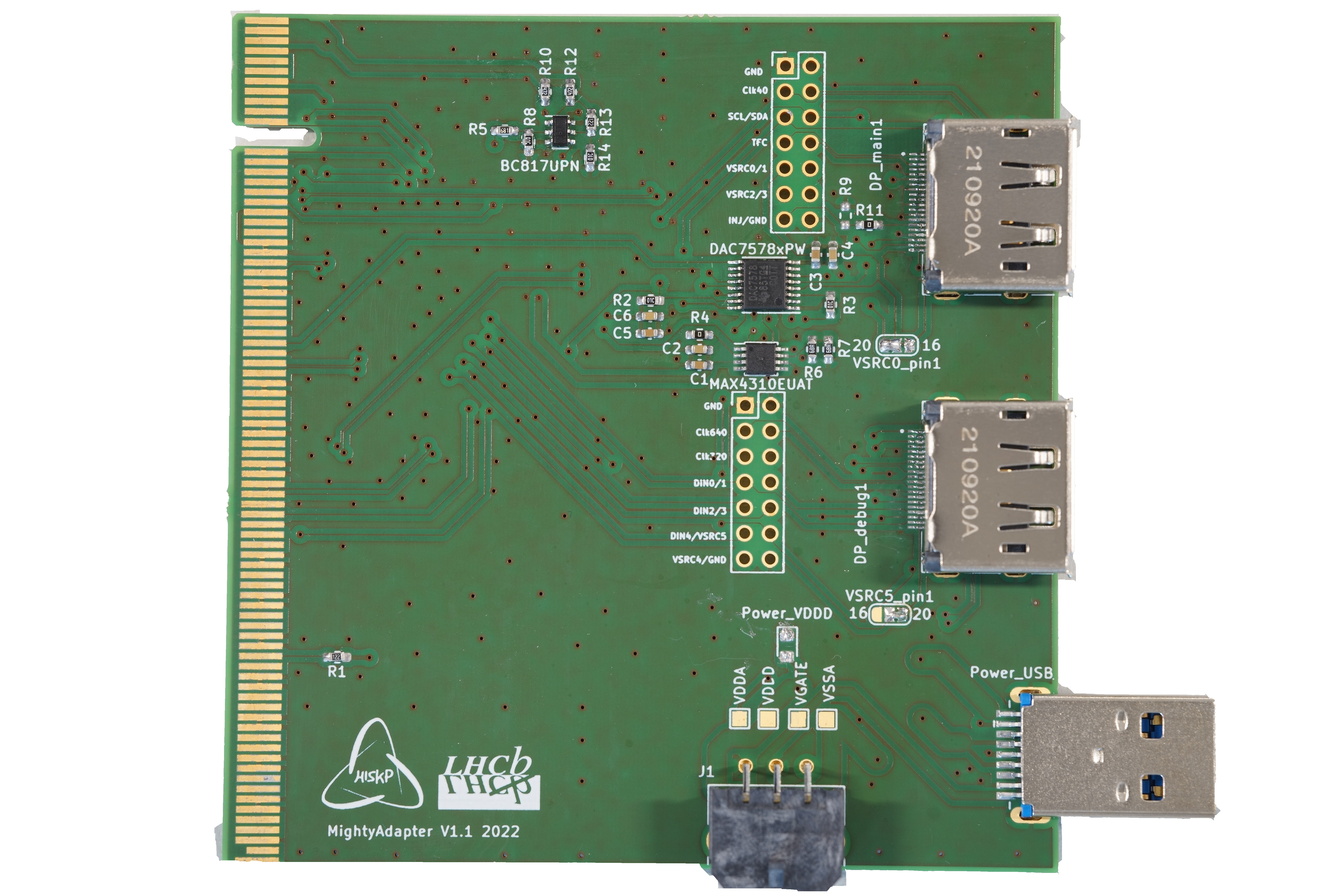}
    \caption{Adapter board equipped with all components and connectors used to connect the FPGA via PCIE express and the chip carrier via display port.}
    \label{fig:AdapterBoard}
\end{figure}

MARS consists of three main components: a FPGA placed on a single PCB~(figure~\ref{fig:FPGABoard}), a adapter board~(figure~\ref{fig:AdapterBoard}) which connects the chip carrier board and the FPGA and the chip carrier itself. The main usage of the adapter besides its role as connector is to provide test points for different signals, e.g, injection, low voltage, clocks and data lines.


\begin{figure}[htbp]
    \centering
    \includegraphics[width=0.7\linewidth]{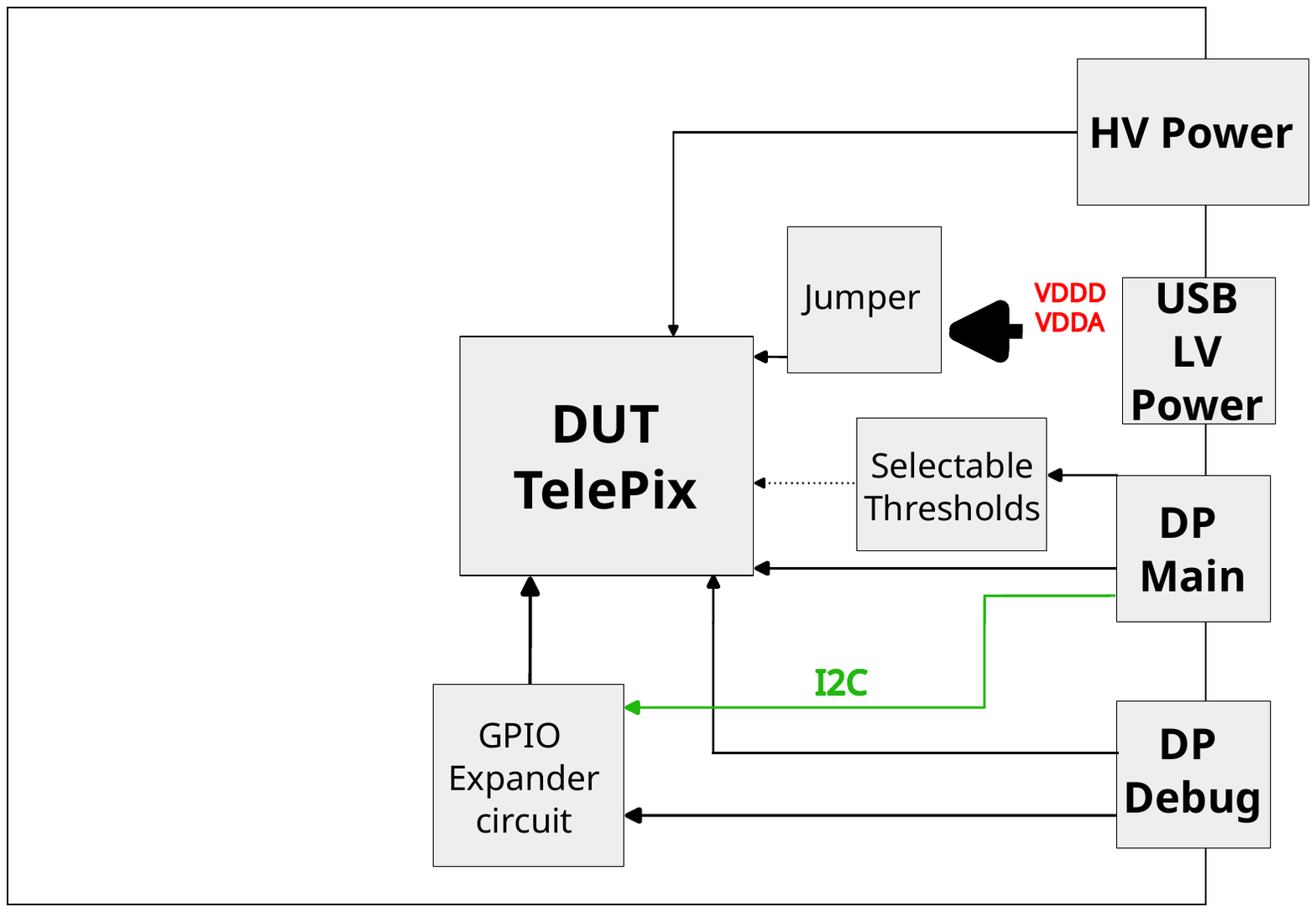}
    \caption{Schematic design of the chip carrier board for the TelePix showing all connectors and main components, e.g., GPIO expander circuit.}
    \label{fig:DUTBoard_schematics}
\end{figure}

The design of the chip carrier boards depends on the chip itself. But for all sensors the overall approach is to keep the amounts of components on the board itself low to provide a good signal quality. An example schematic for the TelePix sensor is shown in figure~\ref{fig:DUTBoard_schematics}.

The main component on the board besides the sensor itself is a GPIO expander which provides more signals for the configuration of the sensor than the hardware intended. The amount of external signal is limited by the display port cables which connects the chip carrier board with the adapter board. Hence, depending on the device components on the chip carrier have to be added.  
But as an advantage the sensitive electronics of the readout system can be placed far away from the chip itself which is crucial at testbeam facilities. 
In addition to the configuration signals for the shift register of the sensor, clock, injection and external voltages for the sensor are provided by the cables. The clocks are either generated by the FPGA or by electronics on the adapter and forwarded by the cables to the carrier board. 

Low voltage can be directly used by a power supply connected via USB to the chip carrier or by an indirect connection via the adapter board.

Besides the development of the hardware components, a new firmware and software was developed based on BASIL~\cite{Basil}. A modular design of the firmware was chosen to allow flexible usage depending on the connected hardware. Additionally, there are different firmware versions depending on the readout speed. This is needed because different hardware connectors of the FPGA have to be addressed depending on the speed. All possible speeds are tested and showed sufficient signal quality.
Moreover, the data quality itself was so far tested for Run2020 and TelePix sensors and shows a good quality at several test points. For the data transfer to the back-end electronics standard Ethernet protocols have been chosen.

The software, used for configuration and characterization, has a python interface to ease the usage. The main design criteria was to keep it modular and flexible. Therefore, each sensor type has a specific class and configuration file, including its specifications, e.g., pixel matrix design and DAC settings. The main functions to program and readout the sensor is similar for all sensors. Hence, new sensors with a similar design can be integrated in the system. Initial tests such as injection tests, threshold scans and overall testing of the signals have been performed for Run2020 and TelePix sensors.

\section{First Results}
\label{sec:Results}

Characterization studies of the first MightyPix prototype are ongoing with the GECCO system~\cite{Lucas}. The transition of the MightyPix to MARS is still ongoing. 

\begin{figure}
    \centering
    \includegraphics[width=0.8\linewidth]{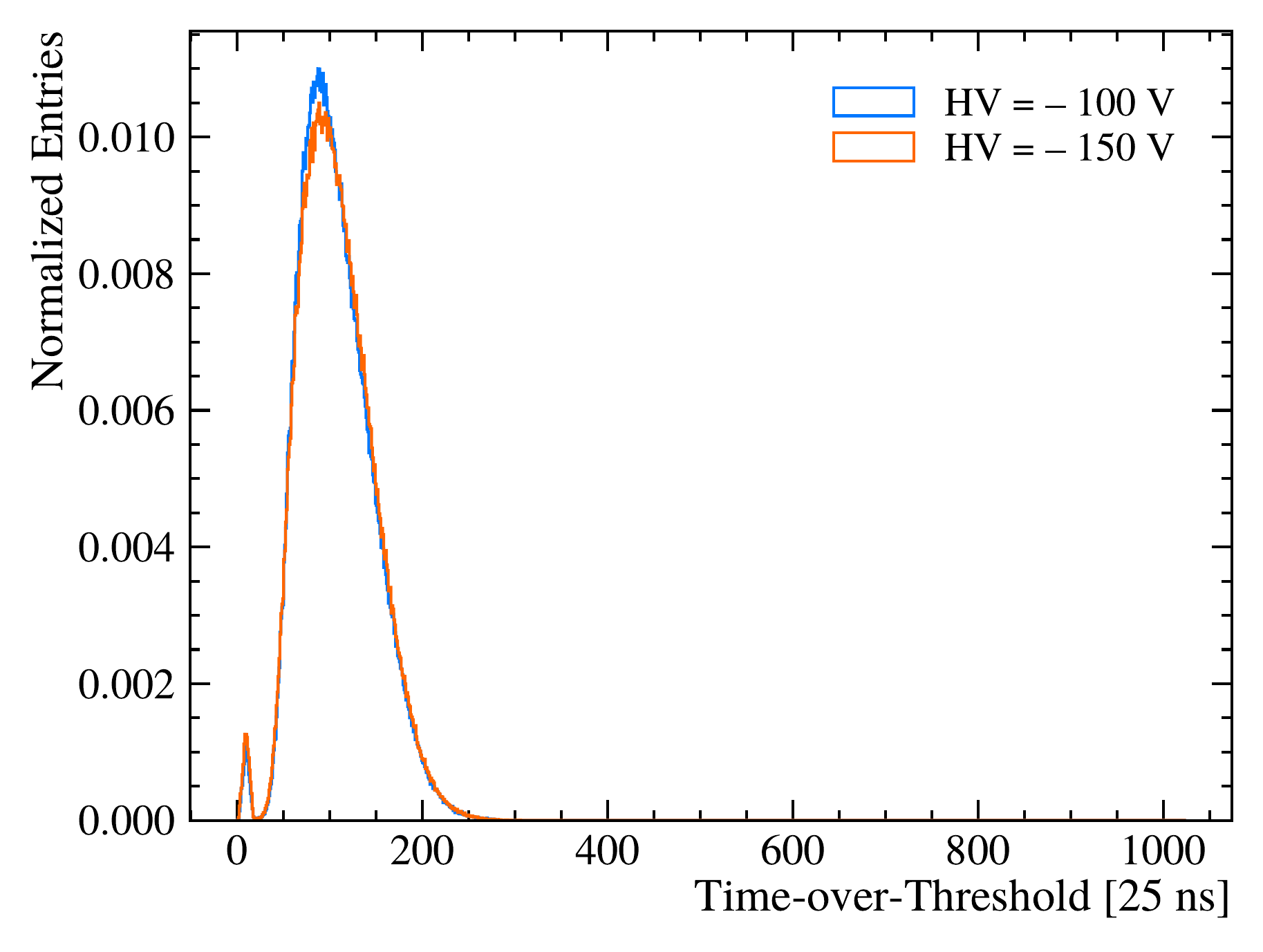}
    \caption{Time-over-Threshold of the MightyPix for a  $^{90}$Sr-source for two different high voltages: $-100~\mathrm{V}$~(colored blue) and $-150~\mathrm{V}$~(colored orange)~\cite{Lucas}.}
    \label{fig:Sr90_MP}
\end{figure}

Main focus of the characterization is the overall performance of the sensor with focus on the power consumption and time resolution. Therefore, the optimal operation settings of the sensor are studied. In addition, different measurements with a $^{90}$Sr-source and a $^{55}$Fe-source were performed to study the sensor performance and Time-over-Threshold ToT. The ToT measurement for different high voltages is shown in figure~\ref{fig:Sr90_MP}. For $-100~\mathrm{V}$ a slightly larger amount of signal compared to $-150~\mathrm{V}$ can be seen. In addition to the mean signal, one small peak can be observed. This is occurred by noise due to line cross-talk between different metal lines in the chip.

Further studies on the time resolution are ongoing. For this a setup consisting of scintillating fibers was developed. Moreover, a testbeam campaign is planned to study the sensor performance further. Noise studies are planned as well.

\section{Summary and Outlook}

A first version of the MightyPix was produced in a $180~\mathrm{nm}$ TSI standard process and successfully tested in the laboratory. First studies show an overall good performance of all implemented electronics and communication protocols. More detailed studies in the laboratory are ongoing. Further studies as the performance in a particle beam and the time resolution as well as the dependence of the performance on the temperature are planned. 

In addition, other development sensors with similar analogue pixel design but different pixel and chip sizes are ongoing. With those a new readout system - MARS - was developed and successfully tested. The integration of the MightyPix into MARS is ongoing. Its performance will first be measured in the laboratory and later in a testbeam campaign as well. Moreover, a proton irradiation of the TelePix at the cyclotron of the University of Bonn is upcoming. Hence, the radiation hardness as well as the influence on the time resolution as well as on the performance will be studied. For this studies MARS will be used.

\bibliographystyle{unsrt}
\bibliography{main}  






\end{document}